\documentclass[twocolumn,twocolappendix]{aastex631}

\shorttitle{CSST void and galaxy clustering surveys}
\shortauthors{Y.Song et al.}
\graphicspath{{./}{figures/}}
\usepackage{subfigure} 
\begin{document}

\title{Cosmological Prediction of the Void and Galaxy Clustering Measurements \\in the CSST Spectroscopic Survey}
\author{Yingxiao Song}
\affiliation{National Astronomical Observatories, Chinese Academy of Sciences,20A Datun Road, Beijing 100012, China}
\affiliation{ School of Astronomy and Space Sciences, University of Chinese Academy of Sciences(UCAS),\\Yuquan Road NO.19A Beijing 100049, China}

\author{Qi Xiong}
\affiliation{National Astronomical Observatories, Chinese Academy of Sciences,20A Datun Road, Beijing 100012, China}
\affiliation{ School of Astronomy and Space Sciences, University of Chinese Academy of Sciences(UCAS),\\Yuquan Road NO.19A Beijing 100049, China}

\author{Yan Gong*}
\affiliation{National Astronomical Observatories, Chinese Academy of Sciences,20A Datun Road, Beijing 100012, China}
\affiliation{ School of Astronomy and Space Sciences, University of Chinese Academy of Sciences(UCAS),\\Yuquan Road NO.19A Beijing 100049, China}
\affiliation{Science Center for China Space Station Telescope, National Astronomical Observatories, Chinese Academy of Sciences,\\20A Datun Road, Beijing 100101, China}

\author{Furen Deng}
\affiliation{National Astronomical Observatories, Chinese Academy of Sciences,20A Datun Road, Beijing 100012, China}
\affiliation{ School of Astronomy and Space Sciences, University of Chinese Academy of Sciences(UCAS),\\Yuquan Road NO.19A Beijing 100049, China}

\author{Kwan Chuen Chan}
\affiliation{School of Physics and Astronomy, Sun Yat-sen University, 2 Daxue Road, Tangjia, Zhuhai, 519082, China}
\affiliation{CSST Science Center for the Guangdong-Hongkong-Macau Greater Bay Area, SYSU, Zhuhai, 519082, China}

\author{Xuelei Chen}
\affiliation{National Astronomical Observatories, Chinese Academy of Sciences,20A Datun Road, Beijing 100012, China}
\affiliation{ School of Astronomy and Space Sciences, University of Chinese Academy of Sciences(UCAS),\\Yuquan Road NO.19A Beijing 100049, China}
\affiliation{Department of Physics, College of Sciences, Northeastern University, Shenyang 110819, China}
\affiliation{Centre for High Energy Physics, Peking University, Beijing 100871, China}

\author{Qi Guo}
\affiliation{National Astronomical Observatories, Chinese Academy of Sciences,20A Datun Road, Beijing 100012, China}
\affiliation{ School of Astronomy and Space Sciences, University of Chinese Academy of Sciences(UCAS),\\Yuquan Road NO.19A Beijing 100049, China}

\author{Guoliang Li}
\affiliation{Purple Mountain Observatory, Chinese Academy of Sciences, Nanjing, 210023, PR China}

\author{Ming Li}
\affiliation{National Astronomical Observatories, Chinese Academy of Sciences,20A Datun Road, Beijing 100012, China}

\author{Yun Liu}
\affiliation{National Astronomical Observatories, Chinese Academy of Sciences,20A Datun Road, Beijing 100012, China}
\affiliation{ School of Astronomy and Space Sciences, University of Chinese Academy of Sciences(UCAS),\\Yuquan Road NO.19A Beijing 100049, China}

\author{Yu Luo}
\affiliation{Purple Mountain Observatory, Chinese Academy of Sciences, Nanjing, 210023, PR China}
\affiliation{Department of Physics, School of Physics and Electronics, Hunan Normal University, Changsha 410081, PR China}

\author{Wenxiang Pei}
\affiliation{National Astronomical Observatories, Chinese Academy of Sciences,20A Datun Road, Beijing 100012, China}
\affiliation{ School of Astronomy and Space Sciences, University of Chinese Academy of Sciences(UCAS),\\Yuquan Road NO.19A Beijing 100049, China}

\author{Chengliang Wei}
\affiliation{Purple Mountain Observatory, Chinese Academy of Sciences, Nanjing, 210023, PR China}

\correspondingauthor{Yan Gong}
\email{Email: gongyan@bao.ac.cn}
 
\begin{abstract}
The void power spectrum is related to the clustering of low-density regions in the large-scale structure (LSS) of the Universe, and can be used as an effective cosmological probe to extract the information of the LSS. We generate the galaxy mock catalogs from Jiutian simulation, and identify voids using the watershed algorithm for studying the cosmological constraint strength of the China Space Station Telescope (CSST) spectroscopic survey. The galaxy and void auto power spectra and void-galaxy cross power spectra at $z=0.3$, 0.6, and 0.9 are derived from the mock catalogs. To fit the full power spectra, we propose to use the void average effective radius at a given redshift to simplify the theoretical model, and adopt the Markov Chain Monte Carlo (MCMC) technique to implement the constraints on the cosmological and void parameters. The systematical parameters, such as galaxy and void biases, and noise terms in the power spectra are also included in the fitting process. We find that our theoretical model can correctly extract the cosmological information from the galaxy and void power spectra, which demonstrates its feasibility and effectivity. The joint constraint accuracy of the cosmological parameters can be improved by $\sim20\%$ compared to that from the galaxy power spectrum only. The fitting results of the void density profile and systematical parameters are also well-constrained and consistent with the expectation. This indicates that the void clustering measurement can be an effective complement to the galaxy clustering probe, especially for the next-generation galaxy surveys.
\end{abstract}


\section {introduction} \label{sec:intro}

On the cosmic scale, the observable universe has a distribution of matter with a network-like structure \citep{1996Natur.380..603B}, and there are obvious filamentary structures connecting them between the galaxy clusters. In the space between these superstructures, galaxy spectroscopic surveys find some regions with large volume and low galaxy density, which are called cosmic voids \citep{1978ApJ...222..784G,1978MNRAS.185..357J,1981ApJ...248L..57K,1986ApJ...302L...1D}. Nowadays, as extensive galaxy surveys provide sufficient galaxy samples in vast space volumes, cosmic voids have become an effective cosmological probe in the studies of cosmic large-scale structure (LSS), such as the void abundance \citep[e.g.][]{contarini2021cosmic,2022A&A...667A.162C,contarini2023cosmological,2023MNRAS.522..152P,songetalvsf,2024arXiv240114451V}, Alcock-Paczy\'{n}ski effect (AP), redshift space distortions (RSD) \citep[e.g.][]{2019MNRAS.483.3472N,nadathur2020completed,correa2021redshift,correa2022redshift,2022A&A...658A..20H}, and baryonic acoustic oscillations (BAO) \citep[e.g.][]{chan2021volume,forero2022cosmic,khoraminezhad2022cosmic}. Cosmic voids are also used in studying modified gravity and massive neutrinos \citep[e.g.][]{cai2015testing,pisani2015counting,zivick2015using,pollina2016cosmic,achitouv2016testing,sahlen2016cluster,falck2018using,sahlen2018cluster,paillas2019santiago,perico2019cosmic,verza2019void,2019MNRAS.488.4413K,2019JCAP...12..055S,contarini2021cosmic,2022ApJ...935..100K,2023A&A...674A.185M,2023JCAP...12..044V,2023JCAP...08..010V}.

Galaxy spectroscopic surveys can map the LSS by providing information on the positions of galaxies in three dimensions, such as  Baryon Oscillation Spectroscopic Survey \citep[BOSS,][]{2017MNRAS.470.2617A} and Dark Energy Spectroscopic Instrument \citep[DESI,][]{2016arXiv161100036D}. The analysis of such data usually focuses on the auto correlation of galaxies, especially the two-point galaxy correlation function (2PCF) and power spectrum. However, galaxy statistics may only provide a part of the LSS information, and we can expect that the LSS probably can be better studied by taking voids into account. 

The void-galaxy cross-correlation has been studied in a number of relevant works \citep[e.g.][]{2016MNRAS.462.2465C,2019MNRAS.483.3472N,2022MNRAS.516.4307W,2022A&A...658A..20H,2023A&A...677A..78R,2023A&A...674A.185M}, which is proven to be an effective cosmological probe for extracting the information of the LSS. 
In this work, we discuss the auto and cross correlations of voids and galaxies in the China Space Station Telescope \citep[CSST,][]{zhan11,zhan2021csst,gong,2023MNRAS.519.1132M} spectroscopic survey, and propose to use the mean void radius at a given redshift to simplify the theoretical calculation based on the halo model for fitting the full power spectra. 

To check the feasibility of this method, we generate the mock galaxy catalogs from Jiutian simulations, and take into account the CSST survey strategy and instrumental design. Then we identify voids in the galaxy catalogs using Voronoi tessellation and watershed algorithm. The void and galaxy auto and cross power spectra at $z$ = 0.3, 0.6, 0.9 are derived from the void and galaxy mock catalogs. After the theoretical modeling of the power spectra, we perform the constraints on the cosmological and void parameters in our model using the Markov Chain Monte Carlo (MCMC) method. The galaxy and void biases and noise terms are also considered in the analysis.

The paper is organized as follows: in Section \ref{sec:data}, we introduce the creation of galaxy and void mock catalogs of the CSST spectroscopic surveys. In Section \ref{sec:powerspectrum}, we estimate the galaxy, void, and void-galaxy power spectra from the mock catalogs and calculate the corresponding theoretical models. In Section \ref{sec:mcmc}, we discuss the constraint results of the model parameters. We summarize our work in Section \ref{sec:conclusion}.

\section{Mock Catalogs} \label{sec:data}

\subsection{Simulation} \label{sec:sim}

We adopt the high-resolution dark-matter-only Jiutian N-body simulations to serve as the basis for generating the galaxy catalog.
The Jiutian simulation we use covers a volume of 1 ($h^{-1}$Gpc)$^3$ and contains $6144^3$ particles with a mass resolution of $m_{\rm p}$ = $3.72 \times 10^8$ $h^{-1}M_\odot$. There are total 128 snapshot outputs from initial redshift $z_i = 127$ down to redshift $z = 0$. The simulation is performed by running the L-Gadget3 code, and uses the friend-of-friend and subfind algorithm to identify the dark matter halo and substructure \citep{2001NewA....6...79S,2005MNRAS.364.1105S}. The best-fit values of the cosmological parameters from $\it Planck$2018 are set as fiducial values in the simulation, i.e. $h = 0.6766$, $\Omega_{\text{m}} = 0.3111$, $\Omega_{\text{b}} = 0.0490$, $\Omega_\Lambda = 0.6899$, $\sigma_8 = 0.8102$ and $n_{\text{s}} = 0.9665$ \citep{2020A&A...641A...6P} .

Considering the RSD and structure evolution effects, we construct each simulation cube with a few slices based on the snapshot outputs at different redshifts. We choose the line-of-sight (LOS) parallel to the edge of the box, and then splice slice-like halo catalogs together according to their comoving distances. In our mock catalog, we address the evolution effects by tracing the merger tree of each galaxy to find the snapshot with the closest redshift corresponding to the galaxy distance. Our method naturally avoids repetition or omission of galaxies at the boundary of slices, compared to directly slicing and stitching the snapshots by redshift. For a reliable RSD calculation, we do not perform the interpolation when splicing the slices, and it will not affect our result at the scales and accuracies we are interested in. We construct three simulation cubes with the central redshift $z_{\rm c} = [0.3,0.6,0.9]$ to build our galaxy and void mock catalogs.

\subsection{Galaxy mock catalog} \label{sec:gcat}

We study the CSST spectroscopic survey for exploring the void and galaxy clustering measurements. The CSST contains a slitless grating spectrograph with three bands i.e. $\it GU$, $\it GV$, and $\it GI$, and covers the wavelength range 225-1000 nm. It will launch in 2026 and plans to survey 17500 ${\rm deg}^2$ sky area in about ten years. The angular resolution is $\sim0.3''$ with 80\% energy concentration for the spectroscopic survey, and the spectral resolution $R = \lambda/\Delta \lambda$ is better than 200. The magnitude limit can reach $\sim23$ AB mag for 5$\sigma$ point source detection in a band.

We construct the mock galaxy catalog using an updated version of the L-Galaxies semi-analytical model \citep{2005MNRAS.364.1105S, 2006MNRAS.365...11C, 2007MNRAS.375....2D, 2011MNRAS.413..101G}, which includes improvements for handling the disruption of satellite galaxies and the growth of supermassive black holes compared to the version from \cite{henriques2015galaxy}. This new model also includes additional features related to galaxy properties, such as the incorporation of galaxy emission line luminosity produced through post-processing techniques \citep{2024MNRAS.529.4958P}.
We can use these emission lines to derive precise spectroscopic redshift and select galaxies that can be detected by the CSST spectroscopic survey. For each galaxy, the redshift $z$ involves the peculiar motions of the source $z_{\text{pec}}$ and cosmological redshift $z_{\text{cos}}$, and the relation is $1+z = (1+z_{\text{cos}})(1+z_{\text{pec}}) = (1+z_{\text{cos}})(1+v_{\rm pec}/c)$, where $v_{\rm pec}$ is the LOS component of peculiar velocity. Besides, we also assign a redshift uncertainty $\sigma = 0.002$ to each galaxy for counting the accuracy of the CSST slitless spectral calibration.

We select galaxies by the signal-to-noise ratio (SNR), and consider four emission lines to estimate the SNR, i.e. H$\alpha$, H$\beta$, [OIII] and [OII]. Since the region of an emission line is usually small compared to the full size of a galaxy, we simply treat galaxies as point sources in the estimation. For a space telescope, the SNR per spectral resolution unit for a spectroscopic sample can be calculated by \citep{cao2018testing,2022MNRAS.515.5894D}
\begin{equation}
    \text{SNR}=\frac{C_{\text{s}} t_{\text{exp}}\sqrt{N_{\text{exp}}}}{\sqrt{C_{\text{s}} t_{\text{exp}}+N_{\text{pix}}[(B_{\text{sky}}+B_{\text{det}})t_{\text{exp}}+R_{\text{n}}^2]}}\label{eq1},
\end{equation}
where $N_{\text{pix}}=\Delta A/l_{\text{p}}^2$ is the number of detector pixels covered by an object. Here $\Delta A$ is the pixel area on the detector, assumed to be the same for all galaxies for simplicity. $l_{\text{p}} = 0.074''$ is the pixel size, and the point-spread function (PSF) is assumed based on the angular resolution of the CSST spectroscopic survey. $N_{\text{exp}}=4$ is the number of exposures and $t_{\text{exp}}=150\, \rm s$ is the exposure time. $R_{\text{n}}$ = 5 $e^-{\text{s}}^{-1}{\text{pixel}}^{-1}$ is the read noise, and $B_{\text{det}}$ = 0.02 $e^-{\text{s}}^{-1}{\text{pixel}}^{-1}$ is the dark current of the detector. $B_{\text{sky}}$ is the sky background in $e^-\text{s}^{-1}\text{pixel}^{-1}$, and $C_{\text{s}}$ is the counting rate from a galaxy. We find that $B_{\text{sky}}=0.016$, 0.196, and 0.266 $e^-\text{s}^{-1}\text{pixel}^{-1}$ for $GU$, $GV$ and $GI$ bands, respectively \citep{songetalvsf}.

We select galaxies if $\rm SNR\ge10$ for any emission line of the four lines mentioned above in any spectroscopic band to get the galaxy mock catalog. We find that the number density of galaxies are $\bar n =  1.5\times 10^{-2}, 2.1\times 10^{-3}, 4.6\times 10^{-4}$  $h^3{\rm Mpc}^{-3}$ for the three redshift bins we choose at $z=0.3$, 0.6, 0.9, respectively. This is basically consistent with the previous studies \citep[e.g.][]{gong}.

\subsection{Void mock catalog}
\label{sec:vcat}

We identify voids in our mock galaxy catalog using Void IDentification and Examination toolkit\footnote{\url{https://bitbucket.org/cosmicvoids/vide\_public/src/master/}} \citep[\texttt{VIDE},][]{vide}, which is based on ZOnes Bordering On Voidness \citep[\texttt{ZOBOV},][]{zobov}. This code identifies voids by Voronoi tessellation and using the watershed algorithm \citep{watershed}, which finds voids with more natural shapes without any shape assumption. It also can provide useful void information, such as void effective radius and volume-weighted center. Note that we use the low-density zones that have not merged before to avoid the void-in-void case when generating the void catalog. 

The voids identified by \texttt{VIDE} are composed of cells containing a galaxy inside, and the total volume $V$ of each void is the sum of all the cell volumes it contains. Based on the volume and position of each cell we can get the void effective radius $R_{\text{v}}$ from $V$, and the volume-weighted center of the void $\mathbf{X}_{\text{v}}$ can be estimated by
\begin{equation}\label{eq2}
\mathbf{X}_{\text{v}} = \frac{1}{V}\sum^N_i\mathbf{x}_i V^i_{\rm cell}.
\end{equation}
Here $V=\sum V^i_{\rm cell}=4/3\pi R_{\rm v}^3$, $V_{\text{cell}}^i$ is the volume of a cell, and $\mathbf{x}_i$ is the coordinate of the galaxy within a cell in a given void.

In Table~\ref{tab:1}, we show the number densities of galaxies and voids in different redshift bins from the mock catalogs. We filter out voids with the effective radius $R_{\rm v}<5\ h^{-1}\text{Mpc}$ to avoid the effects of nonlinear evolution \citep{2021MNRAS.500.4173S}. We also show the average, minimum, and maximum radius of voids with $R_{\rm v}>5\ h^{-1}\text{Mpc}$. We can find that the number density of voids has a similar trend of the galaxy number density, which decreases from $z=0.3$ to 0.9, and the mean void radius becomes larger and larger as redshift increases.

\begin{deluxetable}{cccccc}[h]
\tablenum{1}
\tablecaption{The number densities of galaxy and void (with $R_{\rm v}>$ 5 $h^{-1}\text{Mpc}$), i.e. $n_{\rm g}$ and $n_{\rm v}$ (in $h^3\text{Mpc}^{-3}$), in our mock catalogs at $z=0.3$, 0.6, 0.9. The mean, minimum, and maximum radius of voids (in $h^{-1}\text{Mpc}$) are also shown.\label{tab:1}}
\tablehead{
\colhead{$z$} & \colhead{$n_{\rm g}$} &\colhead{$n_{\rm v}$} & \colhead{$R_{\rm v}^{\rm mean}$} & \colhead{$R_{\rm v}^{\text{min}}$} & \colhead{$R_{\rm v}^{\text{max}}$}
}
\startdata
    0.3& $1.5 \times 10^{-2}$& $4.8 \times 10^{-5}$&11.0&5&45\\
    0.6&$2.1 \times 10^{-3}$& $1.0 \times 10^{-5}$&19.7&5&70\\
    0.9&$4.6 \times 10^{-4}$& $2.9 \times 10^{-6}$&31.4&6&90\\    
\enddata
\end{deluxetable}

\section{POWER SPECTRUM }
\label{sec:powerspectrum}

The galaxy power spectrum in redshift space $P_{\rm AB}^s(k,\mu)$ for two kinds of tracers A and B can be estimated using the power spectrum in real space $P_{\rm AB}(k)$, and we have
\begin{equation}\label{4}
    P_{\rm AB}^s(k,\mu)=P_{\rm AB}(k)(1+\beta\mu^2)^2\mathcal{D}(k,\mu),
\end{equation}
where the superscript ``s" denotes the redshift space, $k$ is the wavenumber, $\mu=k_{\parallel}/k$ is the cosine of the angle between the line of sight and $k$. $\beta= f/b_{\rm g}$, where $b_{\rm g}$ is the galaxy linear bias and $f$ is the growth rate.  $\mathcal{D}(k,\mu)={\rm exp}[-(k\mu\sigma_{\rm D})^2]$ is the damping term at small scales.
Here $\sigma_{\rm D}^2=\sigma_{\rm R}^2+\sigma_{v}^2$, where $\sigma_{\rm R}=c\sigma_ z/H(z)$ is the smearing factor when the power spectrum at the scales smaller than the spectral resolution in the spectroscopic surveys. And $\sigma_z=(1+z)\sigma_z^0$ \citep{2009MNRAS.394.1775W}, we assume $\sigma_z^0=0.002$ considering the accuracy of the CSST slitless spectral calibration. For the velocity dispersion $\sigma_{v}=\sigma_{v0}(1+z)$ \citep{2004PhRvD..70h3007S,2010PhRvD..82f3522T}, we set $\sigma_{v0}=7\ {\rm Mpc}/h$ for the CSST measured emission-line galaxies \citep{gong}. Note that this damping term does not affect our result significantly, since we mainly focus on the linear regime at large scales. Besides, Equation~(\ref{4}) is a simplified model only available at large scales, and more complicated models can be used to describe the small scales up to $k \simeq 0.2\ {\rm Mpc}^{-1}h$ \citep[e.g.][]{2004PhRvD..70h3007S,2024arXiv240407272M}.

For galaxies as tracers,  the real-space galaxy auto power spectrum at $z$ can be estimated by
\begin{equation}\label{7}
    P_{\rm gg}(k,z) \simeq b_{\rm g}^2 P_{\rm mm}(k,z)+N_{\rm g}(z),
\end{equation}
where $N_{\rm g}$ is the galaxy noise term, including the shot noise and systematics in the CSST slitless spectral calibration \citep{gong}. $P_{\rm mm}(k,z)$ is the matter power spectrum at $z$, and we calculate it using \texttt{CAMB} in this work \citep{camb}.

When voids act as tracers, the void auto power spectrum can be estimated by integrating over the void radius based on the halo model as given by \cite{2014PhRvL.112d1304H}. In a narrow range of $R_{\rm v}$, the void power spectrum in real space at $z$ can be simplified as 
\begin{equation}\label{eq8}
P_{\rm vv}(k,z)\simeq [b_{\rm v}(z)\, u_{\rm v}(k,z)]^2 P_{\rm mm}(k,z)+N_{\rm v}(z).
\end{equation}
Here $b_{\rm v}(z)$ is the void bias, and $N_{\rm v}$ is the noise term dominated by the shot noise for voids. $u_{\rm v}(k)$ is the
void density profile in Fourier space, and it can be obtained from the configuration space by
\begin{equation}\label{9}
u_{\rm v}(k)=\frac{\bar{\rho}}{\delta m}\int_{R_{\rm v}^{\rm min}}^{R_{\rm v}^{\rm max}}{u_{\rm v}(r)\frac{\sin (kr)}{kr}4\pi r^2}dr,
\end{equation}
where $\delta m$ is the void uncompensated mass as a normalization factor, which is calculated by
\begin{equation}\label{10}
\delta m=\bar{\rho}\int_{R_{\rm v}^{\rm min}}^{R_{\rm v}^{\rm max}}{u_{\rm v}(r)4\pi r^2}dr.
\end{equation}
Here $u_{\rm v}(r)$ is the void density profile, which denotes the spherically averaged deviation of the void mass density from the mean matter density of the entire universe. It can be calculated using an empirical formula called HSW profile, and it is given by \citep{2014PhRvL.112y1302H}
\begin{equation}\label{11}
u_{\rm v}(r)=\frac {\rho_{\rm v}(r)} {\bar{\rho}}-1=\delta_{\rm cen} \frac {1-(r /R_{\rm s})^\alpha} {1+(r /R_{\rm v})^\beta},
\end{equation}
where $\rho_{\rm v}$ is the void density, $R_{\rm s}\equiv \gamma R_{\rm v}$ is the scale radius when $\rho_{\rm v}$ = $\bar{\rho}$, $\delta_{\rm cen}$ is the central density contrast, and it can be canceled out in the calculation as shown in Equation~(\ref{9}). $\alpha$, $\beta$, and $\gamma$ denote the inner, and outer slope of the compensation wall around the void, and the ratio of scale radius relative to the void radius, respectively, and we set them as free parameters which can be jointly fitted in the model fitting process.

\begin{figure*}
    \centering
	\includegraphics[width=1.9\columnwidth]{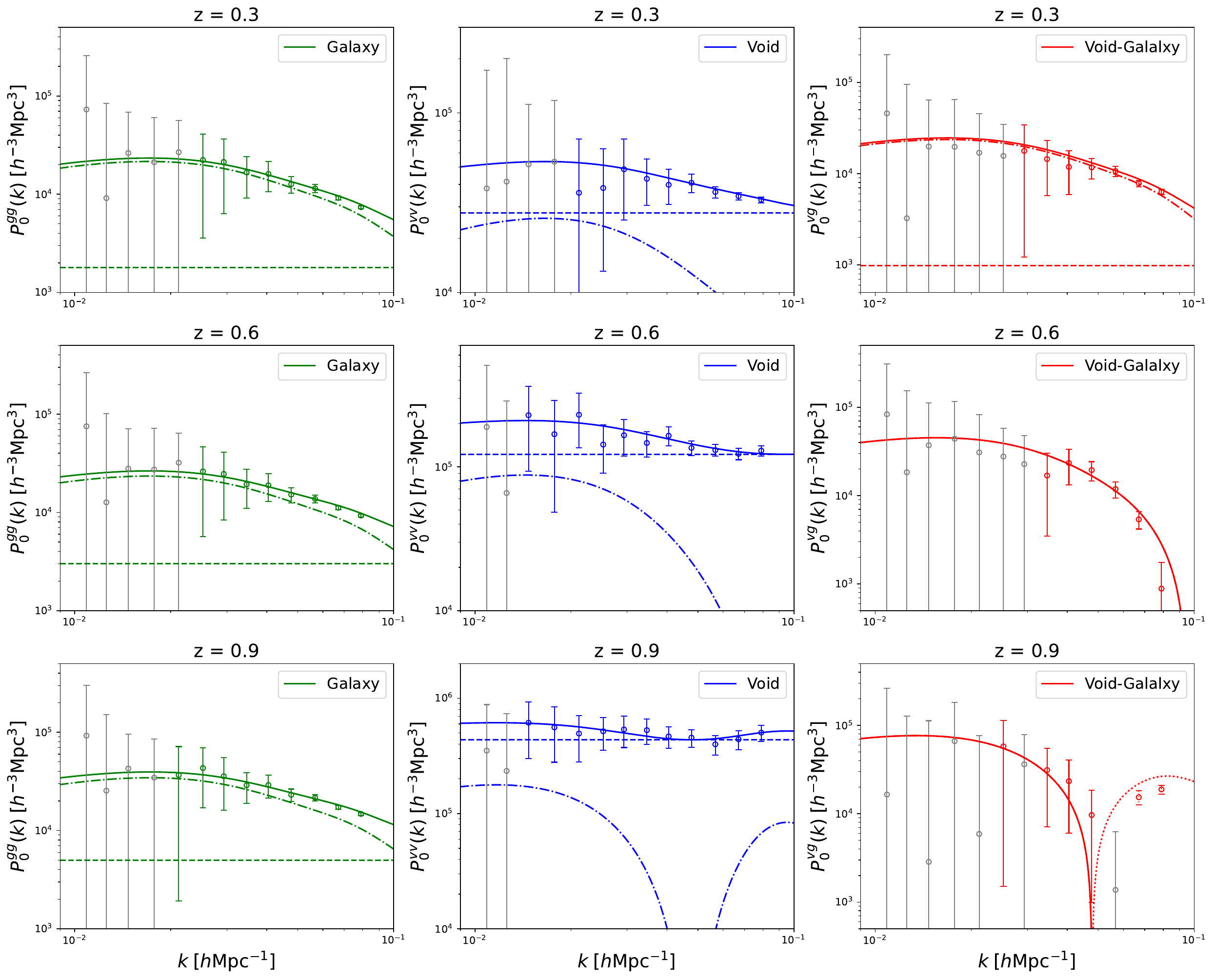}
    \caption{The multipole power spectra $P^{\rm gg}_{\ell}$ (green), $P^{\rm vv}_{\ell}$ (blue) and  $P^{\rm vg}_{\ell}$ (red) with $\ell=0$ at $z=0.3$, 0.6 and 0.9 in the simulation. The gray data points with SNR$<$1 are excluded in the constraint process.  The curves are the best-fits of the theoretical calculation, and the dotted line indicates that the values are negative. The solid, dash-dotted, and dashed curves are the total, clustering, and noise terms of the power spectra.}
    \label{fig:ps}
\end{figure*}

Here we propose to use the mean void effective radius $R_{\rm v}^{\rm mean}$ to simplify the calculation, by setting $R_{\rm v}=R_{\rm v}^{\rm mean}$ in Equation~(\ref{11}). This means $R_{\rm s}$, $u_{\rm v}(r)$ and $u_{\rm v}(k)$, as well as $\alpha$, $\beta$ and $\gamma$, can be seen as the mean values of all selected voids at $z$. Then we can use Equation~(\ref{eq8}) to calculate the void auto power spectrum as a good approximation.

Similarly, we can estimate the void-galaxy cross-power spectrum at $z$ by \citep{2014PhRvL.112d1304H}
\begin{equation}\label{12}
P_{\rm vg}(k)= b_{\rm v} b_{\rm g} u_{\rm v}(k) P_{\rm mm}(k)+N_{\rm v}u_{\rm v}(k)+N_{\rm vg},
\end{equation}
where $N_{\rm vg}$ is the noise term for the cross correlation.
Besides the three parameters in the HSW profile, we also set the galaxy bias $b_{\rm g}$, void bias $b_{\rm v}$, and noise terms of the auto and cross power spectra at a given redshift as free parameters, which can provide proper flexibility to explore the distribution of voids in the LSS at different redshifts.


We obtain the multipole power spectrum by integrating $P_{\rm AB}^s(k,\mu)$ over $\mu$ and considering the Alcock-Paczy\'{n}ski effect \citep[AP,][]{1979Natur.281..358A}\footnote{Note that the AP effect is probably more complicated in the power spectrum modelling, especially for voids \citep[e.g.][]{2024arXiv240702699R}. The current AP effect model may not fully describe the data, and we may need to consider a more complex model of the AP effect in the void and galaxy surveys.}. The multipole power spectrum is given by
\begin{equation}\label{eq6}
    P_{\ell}^{\rm AB}(k)=\frac{2\ell+1}{2\alpha_{\perp}^2\alpha_{\parallel}}\int_{-1}^1P_{\rm AB}^s(k',\mu')\mathcal{L}_{\ell}(\mu)d\mu.
\end{equation}
Here $\mathcal{L}_{\ell}(\mu)$ is the Legendre polynomials, and only the non-vanishing components $\ell=(0,2,4)$ need to be considered. $\alpha_\perp=D_{\rm A}(z)/D_{\rm A}^*(z)$ is the transverse scaling factor and $\alpha_\parallel=H^*(z)/H(z)$ is the radial scaling factor. The superscript ``*" means the fiducial cosmology. The apparent wavenumber and cosine of angle are derived by $k'=\sqrt{{k'_\parallel}^2+{k'_\perp}^2}$ and $\mu'=k'_\parallel/k'$, where $k'_\perp=k_\perp/\alpha_\perp$ and $k'_\parallel=k_\parallel/\alpha_\parallel$. Note that, we only consider the multipole power spectra with $\ell=0$ as an example in this work, since the data for $\ell=2$ and 4 have large errors and low SNRs limited by the simulation box size. We have tested and found that including the data for $\ell=2$ and $\ell=4$ in the current analysis does not significantly improve our constraint results. We can use the full multipole power spectra with $\ell=(0,2,4)$ in the real CSST or other Stage IV surveys with large sky coverage.

For generating the mock data of the void and galaxy power spectra, we make use of \texttt{powerbox} \citep{2018JOSS....3..850M} to derive the data points, and the errors are estimated by using the jackknife method. In order to obtain sufficient statistical significance in the fitting process, we only use the data points with SNR $>$ 1. In Figure \ref{fig:ps}, we show the mock data of the galaxy, void, and void-galaxy multipole power spectra with $\ell=0$ at $z=0.3$, 0.6, and 0.9. We note that, due to the limitation of the size of the simulation box, the data points have quite large errors and variances at large scales with $k\lesssim0.02\ h/{\rm Mpc}$. This issue can be significantly improved in the real CSST or other Stage IV surveys, covering several thousand or more than ten thousand square degrees.  Besides, we only consider the scales at $k<0.1\ {\rm Mpc}^{-1}h$ to avoid the nonlinear effects, which are difficult to accurately model.

\begin{deluxetable*}{ccccccc}
\renewcommand{\arraystretch}{1.25}
\tablenum{2}
\tablecaption{The fiducial values, flat prior, best-fit values and errors of the six cosmological parameters, three void density profile parameters $\alpha^i,\beta^i,\gamma^i$, and the void biases $b_{\rm v}^i$ at $z=0.3$, 0.6, and 0.9. The relative constraint accuracies of the parameters are also shown. Note that these results are derived from the simulations limited by the box size. If considering the 17,500 deg$^2$ full sky coverage of the CSST, the constraint accuracy can be improved by about one order of magnitude. \label{tab:3}}
\tablehead{
\colhead{Parameter} &\colhead{Fiducial value}&
\colhead{Flat prior} & \colhead{Constraints by}&
\colhead{Constraints by}&
\colhead{Joint constraints}\\
&&&\colhead{galaxy clustering}&\colhead{void clustering}&
}
\startdata
Cosmology\\
\hline
$w$&-1&(-1.8, -0.2)&  $-1.117_{-0.472}^{+0.507}(43.8\%)$&$-0.824_{-0.710}^{+0.522}(74.7\%)$&$-1.207_{-0.431}^{+0.444}(36.2\%)$\\
$h$& 0.6766&(0.5, 0.9)&  $0.609_{-0.067}^{+0.094}(13.2\%)$&$0.603_{-0.076}^{+0.166}(20.0\%)$&$0.600_{-0.059}^{+0.066}(10.4\%)$\\
$\Omega_{\text{m}}$& 0.3111&(0.1, 0.5)& $0.362_{-0.090}^{+0.089}(24.6\%)$&$0.313_{-0.151}^{+0.133}(45.3\%)$&$0.371_{-0.082}^{+0.082}(22.1\%)$\\
$\Omega_{\text{b}}$&0.049&(0.02, 0.08) &$0.042_{-0.016}^{+0.021}(44.5\%)$&$0.047_{-0.019}^{+0.022}(43.2\%)$&$0.035_{-0.011}^{+0.021}(45.5\%)$\\
$n_{\text{s}}$&0.9665&(0.7, 1.2)& $0.884_{-0.126}^{+0.156}(15.9\%)$&$1.022_{-0.217}^{+0.132}(17.1\%)$&$0.921_{-0.144}^{+0.168}(17.0\%)$\\
$A_{\text{s}}(\times 10^{-9})$&2.1&(1.0, 3.0)& $2.049_{-0.647}^{+0.622}(30.9\%)$&$1.827_{-0.623}^{+0.770}(38.1\%)$&$1.920_{-0.526}^{+0.654}(30.7\%)$\\
\hline
Void\\
\hline
$\alpha^1$&-&(0, 10.0)&-&$2.978_{-2.135}^{+3.775}(99.2\%)$&$2.216_{-1.579}^{+2.587}(94\%)$\\
$\alpha^2$&-&(0, 10.0)& -&$2.915_{-1.935}^{+3.660}(96.0\%)$&$4.448_{-3.074}^{+3.265}(71.3\%)$\\
$\alpha^3$&-&(0, 10.0)& -&$3.149_{-2.333}^{+3.973}(100.1\%)$&$1.878_{-1.351}^{+2.366}(99.0\%)$\\
$\beta^1$&-&(0, 20.0)&-&$9.501_{-5.832}^{+5.649}(60.4\%)$&$11.275_{-3.784}^{+4.867}(38.4\%)$\\
$\beta^2$&-&(0, 20.0)& -&$10.700_{-4.149}^{+5.519}(45.2\%)$&$11.893_{-2.890}^{+3.510}(26.9\%)$\\
$\beta^3$&-&(0, 20.0)& -&$14.657_{-4.823}^{+3.239}(27.5\%)$&$14.122_{-2.825}^{+2.077}(17.4\%)$\\
$\gamma^1$&-&(0, 2.0)& -&$0.706_{-0.516}^{+0.824}(94.9\%)$&$0.438_{-0.301}^{+0.418}(82.0\%)$\\
$\gamma^2$&-&(0, 2.0)& -&$0.709_{-0.506}^{+0.792}(91.5\%)$&$0.854_{-0.070}^{+0.049}(7.0\%)$\\
$\gamma^3$&-&(0, 2.0)& -&$0.958_{-0.635}^{+0.728}(71.2\%)$&$0.795_{-0.046}^{+0.057}(6.5\%)$\\
$b_{\rm v}^1$&-&(-10, 10)& -&$1.142_{-1.676}^{+1.071}(120.2\%)$&$1.380_{-0.352}^{+0.374}(26.3\%)$\\
$b_{\rm v}^2$&-&(-10, 10)& -&$1.932_{-3.504}^{+2.136}(145.9\%)$&$3.185_{-0.701}^{+0.869}(24.6\%)$\\
$b_{\rm v}^3$&-&(-10, 10)& -&$2.595_{-4.406}^{+3.282}(148.1\%)$&$5.724_{-1.534}^{+1.857}(29.6\%)$\\
\enddata
\end{deluxetable*}

\section{CONSTRAINT AND RESULTS}
\label{sec:mcmc}

\begin{figure*}
    \centering
	\includegraphics[width=1.9\columnwidth]{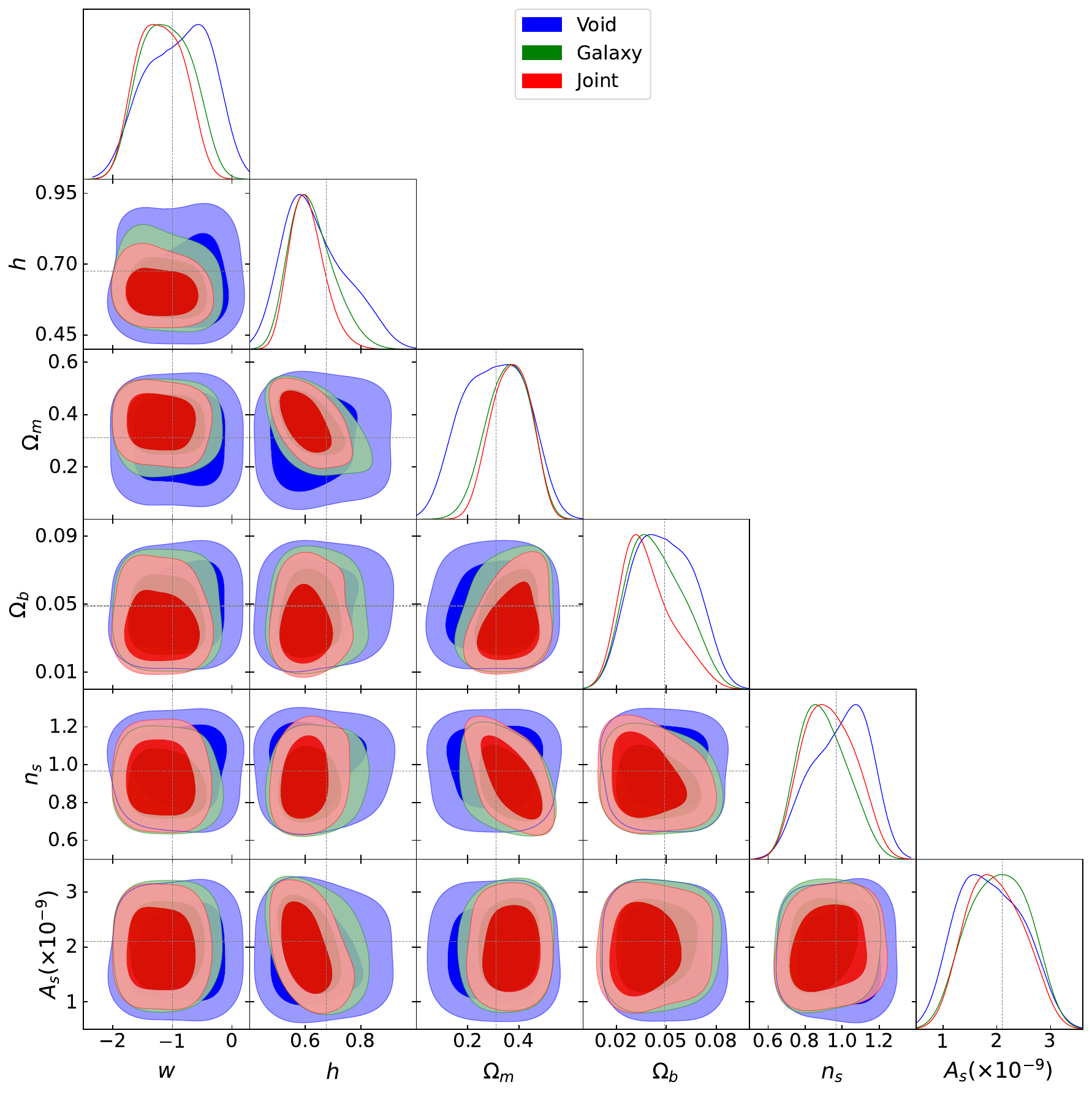}
    \caption{Contour maps of the six cosmological parameters at 68\% and 95\% CL for galaxy clustering (green), void clustering (blue), and joint constraints (red) derived from the simulation. The 1D PDF for each parameter is also shown. The gray dotted lines mark the fiducial values of the cosmological parameters.}
    \label{fig:mcmccosmic}
\end{figure*}

\begin{figure*}
\centering
\subfigure{ \includegraphics[width=0.66\columnwidth]{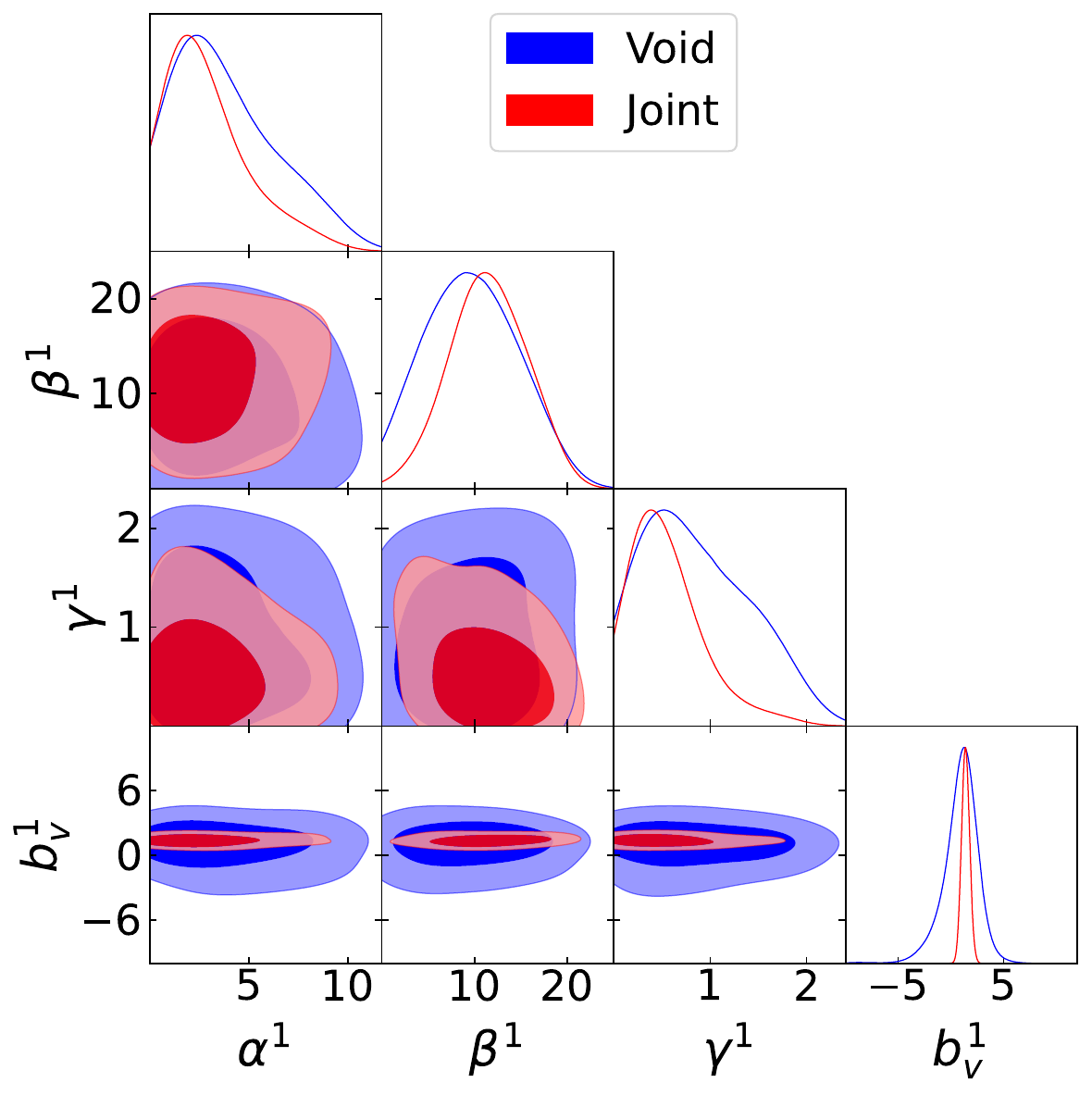}}
\subfigure{ \includegraphics[width=0.66\columnwidth]{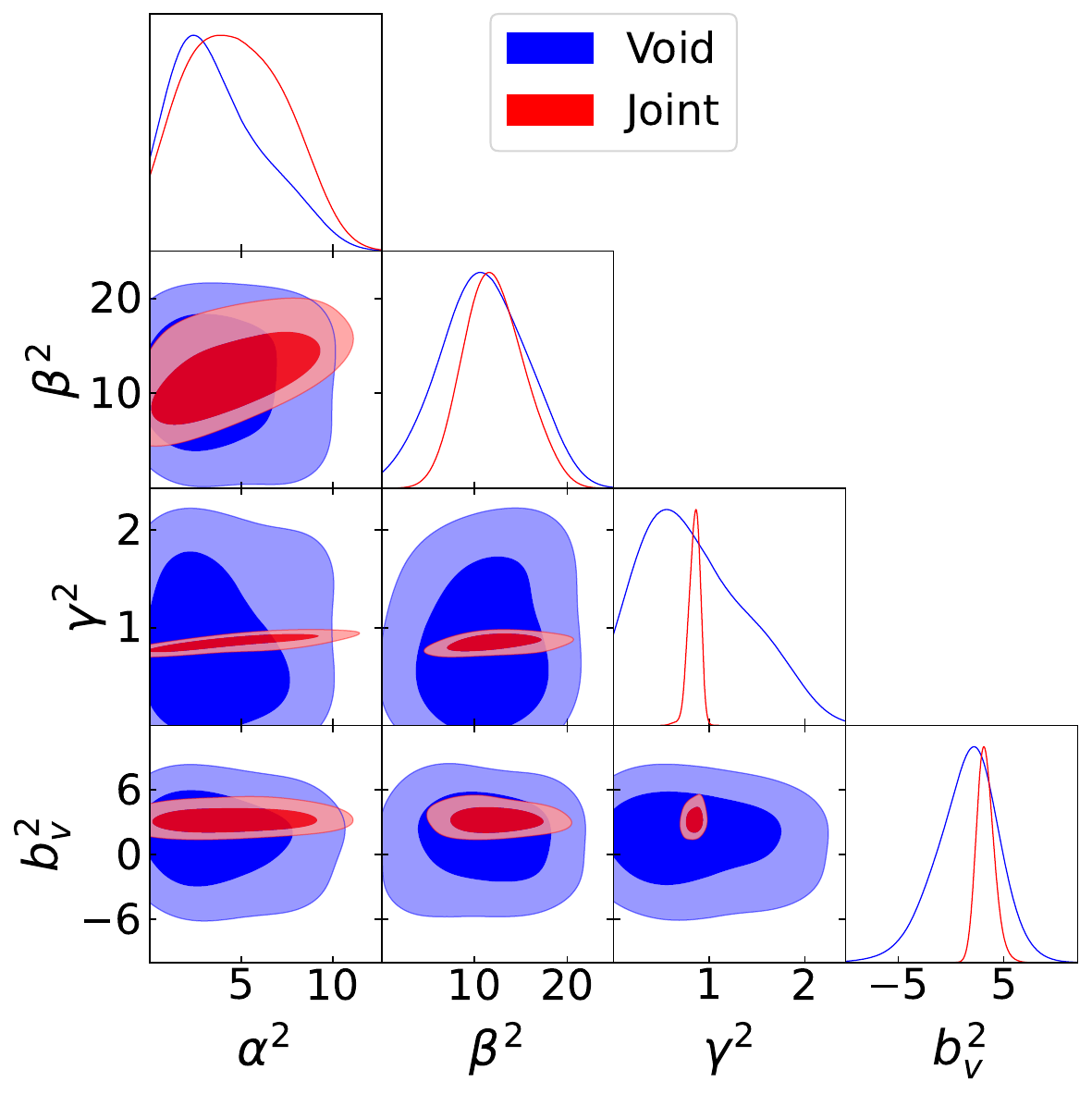}}
\subfigure{ \includegraphics[width=0.66\columnwidth]{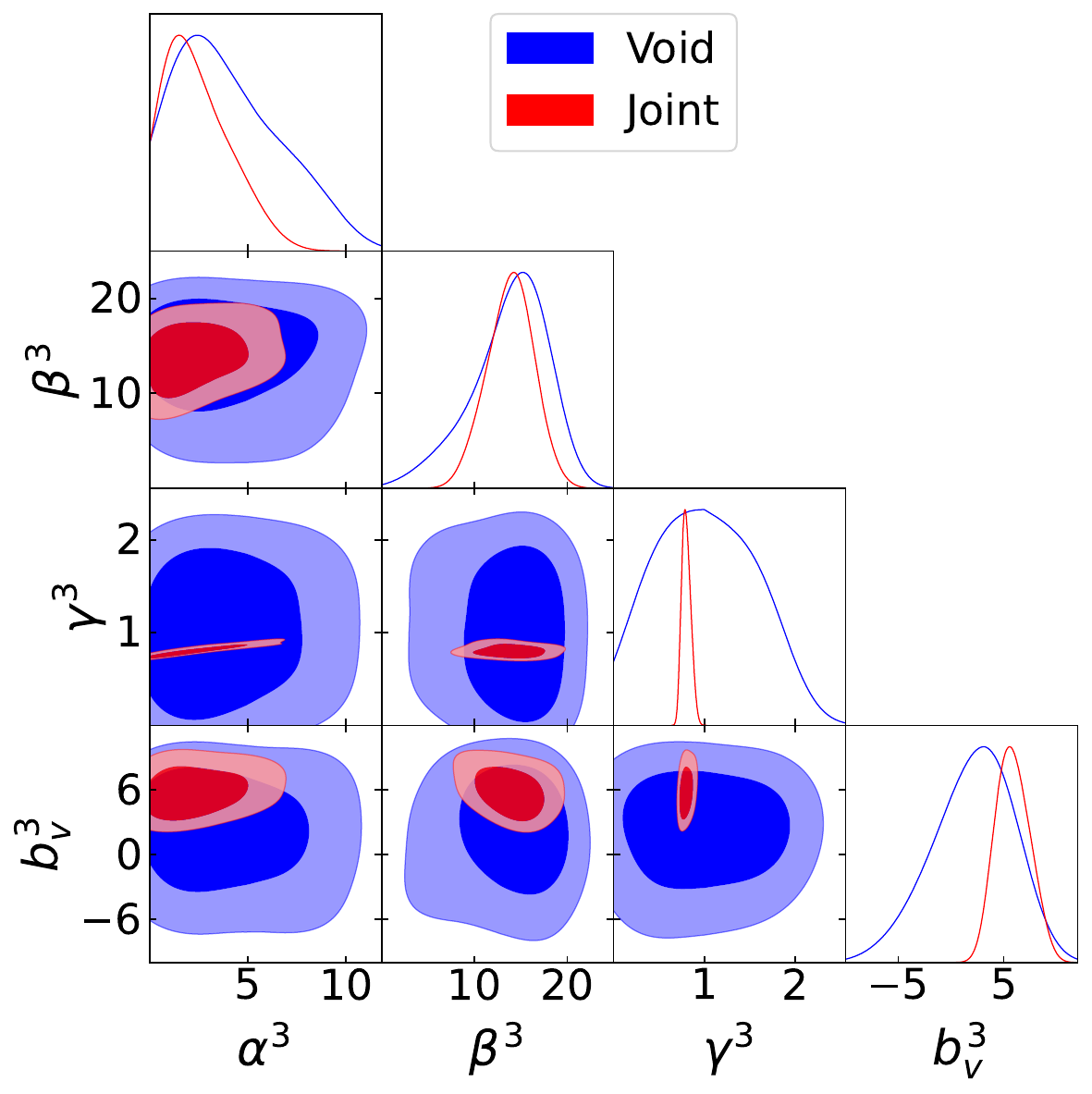}}
\caption{Contour maps at 68\% and 95\% CL and 1D PDFs of the three void density profile parameters and void biases constrained by the void auto power spectra (blue) and joint power spectra (red) derived from the simulation. The left, middle and right panels show the results of $\alpha^i$, $\beta^i$, $\gamma^i$ and $b_{\rm v}^i$ at $z=0.3$, 0.6 and 0.9, respectively.}
\label{fig:mcmcvoid}
\end{figure*}

The $\chi^2$ method is adopted to fit the mock data of void, galaxy and void-galaxy power spectra, which takes the form as
\begin{equation}\label{eq13}
    \chi_{\rm AB}^2 = \sum\frac{\left[P^{\rm AB}_{\rm data}(k,z)-P^{\rm AB}_{\rm th}(k,z)\right]^2}{\sigma_{\rm data}^2},
\end{equation}
where $P^{\rm AB}_{\rm data}(k,z)$ and $P^{\rm AB}_{\rm th}(k,z)$ are the mock data and the theoretical power spectrum at $z$, respectively, and $\sigma_{\rm data}$ is the error of the mock data. Then the likelihood function can be estimated by $\mathcal{L}$ $\propto$ exp($-\chi^2$/2). The total chi-square for the joint constraint can be estimated by $\chi^2_{\rm joint} = \chi^2_{\rm gg} + \chi^2_{\rm vv} + \chi^2_{\rm vg}$, where $\chi^2_{\rm gg}, \chi^2_{\rm vv}$ and $\chi^2_{\rm vg}$ are the chi-squares for the galaxy, void and void-galaxy power spectra. Note that, for simplicity, we do not consider the correlations of the power spectra in our analysis, i.e. the covariance matrix, which could make the constraints worse in the joint constraint. We will estimate and test the effect of the covariance matrix between the galaxy, void, and void-galaxy power spectra using simulations in our future work.

We constrain the free parameters in our fitting process using the MCMC method by \texttt{emcee} \citep{emcee,goodman}. We choose 112 walkers and obtain 30000 steps for each walker. The first 3000 steps have been removed as the burn-in process, and after thinning the chains, we obtain 30,000 chain points for illustrating the probability distribution functions (PDFs) of the free parameters. 
In Table \ref{tab:3}, we list the fiducial values and flat priors for the cosmological and void parameters. 
The free cosmological parameters include dark energy equation of state  $w$, reduced Hubble constant $h$, spectral index $n_{\rm s}$, and amplitude of initial power spectrum $A_{\rm s}$, the total matter density parameter $\Omega_{\text{m}}$, and baryon density parameter $\Omega_{\rm b}$. Note that the fiducial value of $A_{\rm s}$ is derived by $\sigma_8=0.8102$ from the Jiutian simulation. Note that, since the free parameters about void density profile, i.e. $\alpha$, $\beta$, and $\gamma$, are not the input parameters in the simulation, they do not have the fiducial values.

Besides, the flat priors of the systematical parameters are $b_{\rm g}^i \in (0,3.0)$, $b_{\rm v}^i \in (-10.0,10.0)$, ${\rm log}_{10}(N^i_{\rm g}) \in (3.0,5.0)$, ${\rm log}_{10}(N^i_{\rm v}) \in (0,10.0)$ and ${\rm log}_{10}(N^i_{\rm vg}) \in (-10.0,10.0)$ at a given redshift. In the joint constraint, to obtain better constraint results, we consider the fitting results from the galaxy auto power spectra, and set tighter prior ranges for $N_{\rm g}^i$ with the lower bounds $1 \times 10^3, 2 \times 10^3$ and $3 \times 10^3 ({\rm Mpc}/h)^{-3}$ at the three redshifts. In our fitting process, we totally have 30 free parameters, which contain 6 cosmological parameters, 3 void parameters at each redshift, and 15 systematical parameters.

In Figure~\ref{fig:mcmccosmic}, we show the constraint results of the cosmological parameters from galaxy clustering only, void power spectrum only, and joint constraint. The best-fit values, 1$\sigma$ errors, and relative accuracies for the cosmological parameters are listed in Table \ref{tab:3}. We find that the minimum reduced chi-square $\chi^2_{\rm red}$ from the joint fitting process is smaller than 1, which indicates our model is effective in fitting the data of the multipole power spectra with $\ell=0$. Note that the current constraint accuracies are derived from the simulations limited by the box size, and it will be improved by about one order of magnitude for the real CSST full sky survey with 17,500 deg$^2$. The best-fit curves of the void, galaxy, and void-galaxy power spectra, and the clustering and noise terms are shown in Figure~\ref{fig:ps}. 

We can find that the best-fit curves can match the data points of the power spectra very well (see Figure~\ref{fig:ps}), and the fitting results of the cosmological parameters are consistent with the fiducial values within or close to 1$\sigma$ confidence level (CL) (see Figure~\ref{fig:mcmccosmic}). This means that our theoretical model can explain the data and extract the cosmological information correctly, especially for the modeling of the void auto power spectrum and void-galaxy cross-power spectrum. 

We also notice that, as shown in Figure~\ref{fig:ps}, the void power spectrum is actually dominated by the noise term, particularly at high redshifts, due to large shot noise or low number density, which may limit its constraint accuracy on the cosmological parameters. However, after considering the void-galaxy power spectrum, the joint fitting can improve the constraint accuracies by as large as $\sim20\%$ compared to that from the galaxy power spectrum only\footnote{Note that we can include the data at smaller scales if choosing a more accurate and complicated model of the galaxy power spectrum (as shown by Equation~(\ref{4})). This will improve the constraint power of the galaxy power spectrum, and may suppress the gain of including voids in the joint fitting process.}. This indicates that the probe of void and void-galaxy power spectra can be complementary to the galaxy clustering measurement for extracting cosmological information, especially for the next generation of surveys, e.g. CSST, covering large sky areas with high magnitude limits.

In Figure \ref{fig:mcmcvoid}, we show the contour maps and 1D PDFs of $\alpha$, $\beta$, $\gamma$ and $b_{\rm v}$ derived from the void power spectrum only and joint constraint. We can find that the best-fit values of $\alpha$ are similar at the three redshifts, which give $\alpha\sim3$. For $\beta$ and $\gamma$, it has some trend that both of them become larger at higher redshifts, which varies from 10 to 15 for $\beta$ and from 0.4 to 0.9 for $\gamma$, although the best-fit values are consistent within 1$\sigma$ at the three redshifts. This is also consistent with expectations, that the density profile of voids becomes flatter and shallower at high redshifts. Besides, the joint constraints can effectively improve the constraint accuracies of the void parameters, especially for $\gamma$, which is because the cosmological and systematical parameters can be better constrained in this case. These results can provide reference and be significantly improved in future galaxy surveys.

For the constraint on $b_{\rm v}$, we note that the joint power spectra can effectively improve the constraint accuracy by several times or even one order of magnitude, compared to the case using the void auto power spectra only (see also Table~\ref{tab:3}). We can also find that the probability distributions of $b_{\rm v}$ at the three redshifts cover both positive and negative values while the positions of the peaks are positive, if only considering the void auto power spectrum. On the other hand, the probability distributions of $b_{\rm v}$ are mainly in the positive regions if using the joint power spectra. This is because, since the void bias is usually negative at large scales and positive at small scales \citep[see e.g.][]{2014PhRvL.112d1304H,chan2014large}, and our mock data at small scales with much smaller errors are dominant (see Figure~\ref{fig:ps}), especially for the void-galaxy power spectra, the current results are reasonable and expectable.

Besides the void bias $b_{\rm v}$, the other systematical parameters in the theoretical models of the galaxy power spectrum, void power spectrum and void-galaxy power spectrum are also jointly constrained, such as the galaxy bias $b_{\rm g}$, and noise terms $N_{\rm g}$, $N_{\rm v}$ and $N_{\rm vg}$ at different redshifts. The constraint results of these parameters are shown in the Appendix, and we can find that these parameters are also well-constrained in our analysis.


\section{Summary and conclusion}
\label{sec:conclusion}

In this work, we study the constraints on the cosmological and void parameters using the void and galaxy clustering mock data measured by the CSST spectroscopic survey. The Jiutian simulation is used to obtain the galaxy mock catalog at $z=0.3$, 0.6, and 0.9, considering the CSST instrumental design and survey strategy. We identify voids from the galaxy catalogs by \texttt{VIDE} which adopts the watershed algorithm, and generate the void mock catalogs. Then the void power spectrum, galaxy power spectrum, and void-galaxy cross-power spectrum at each redshift are derived. In order to fit the data, we propose to use the mean void effective radius $R_{\rm v}^{\rm mean}$ at a given redshift to simplify the theoretical model. The MCMC method is employed to perform the constraint, and the systematical parameters are also considered in the fitting process.

We find that our theoretical model can fit the galaxy, void, and void-galaxy power spectra very well, and the best fits of the cosmological parameters are consistent with their fiducial values within or close to 1$\sigma$ CL, which demonstrates the feasibility of our modeling. Besides, the joint constraints including all galaxy and void power spectra can effectively improve the constraint accuracy of the cosmological parameters by $\sim20\%$, compared to the result derived from the galaxy power spectrum only. This indicates that the void clustering measurement can be a good complement to the galaxy clustering probe, especially for future wide-field galaxy surveys with large survey volumes and massive samples. 
For the parameters of the void density profile, i.e. $\alpha$, $\beta$, and $\gamma$, we also obtain stringent constraint results in the joint fitting case, and find that the void density profile becomes flatter and shallower at high redshifts, which is as expected for the properties of high-$z$ voids. 

Considering the full CSST survey area with 17500 deg$^2$, it is expected that more accurate data on the larger scales at  $k\lesssim0.01\ {\rm Mpc}^{-1}h$ can be obtained, and the void and galaxy clustering measurements from the real CSST surveys can improve the constraints on the cosmological and void parameters to be the order of a few percent or even higher accuracy. This means that cosmic void observations have great potential in the next-generation galaxy surveys, and can take an important role in future cosmological studies.

\begin{acknowledgments}
YS, QX and YG acknowledge the support from National Key R\&D Program of China grant Nos. 2022YFF0503404, 2020SKA0110402, and the CAS Project for Young Scientists in Basic Research (No. YSBR-092). KCC acknowledges the support the National Science Foundation of China under the grant number 12273121. XLC acknowledges the support of the National Natural Science Foundation of China through Grant Nos. 11473044 and 11973047, and the Chinese Academy of Science grants ZDKYYQ20200008, QYZDJ-SSW-SLH017, XDB 23040100, and XDA15020200. QG acknowledges the support from the National Natural Science Foundation of China (NSFC No.12033008). GLL, YL and CLW acknowledges the support from NSFC grant No. U1931210. This work is also supported by science research grants from the China Manned Space Project with Grant Nos. CMS- CSST-2021-B01 and CMS-CSST-2021-A01. 
\end{acknowledgments}

\vspace{3mm}

\appendix

\begin{figure}
\centering
\includegraphics[width=\columnwidth]{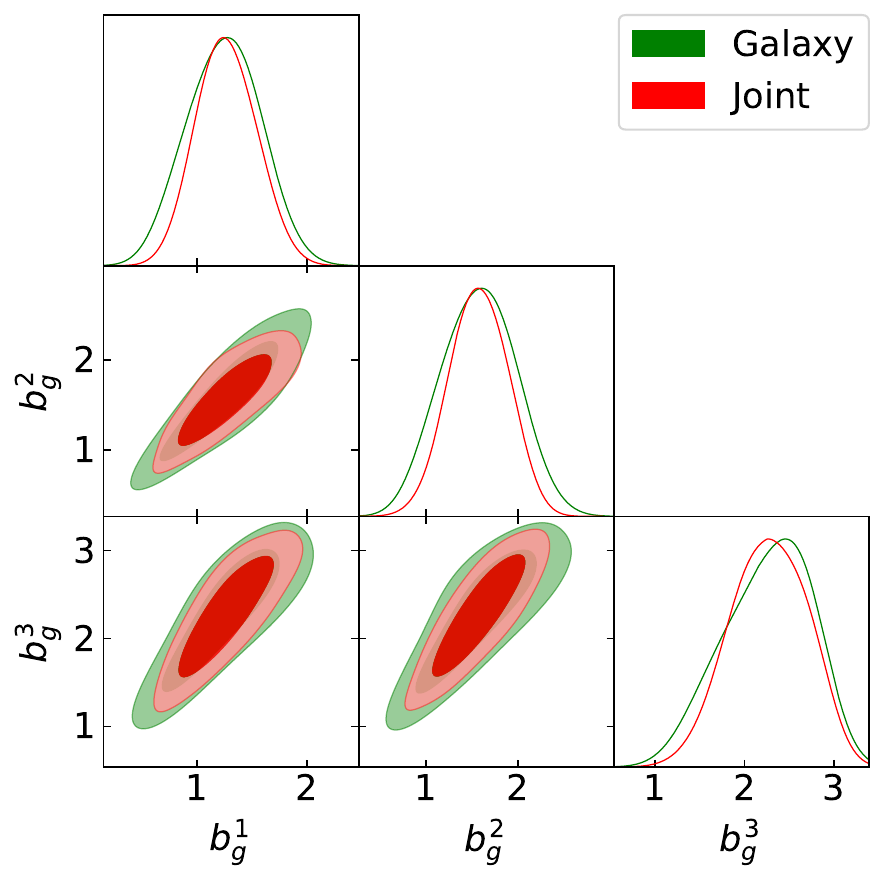}

\caption{Contour maps at 68\% and 95\% CL and 1D PDFs of the galaxy bias $b^i_{\rm g}$ at $z=0.3$, 0.6 and 0.9 derived from the galaxy power spectrum (green) and joint power spectra (red) in the simulation, respectively.}
\label{fig:mcmcbias}
\end{figure}

\begin{figure*}
\centering
\subfigure{ \includegraphics[width=0.66\columnwidth]{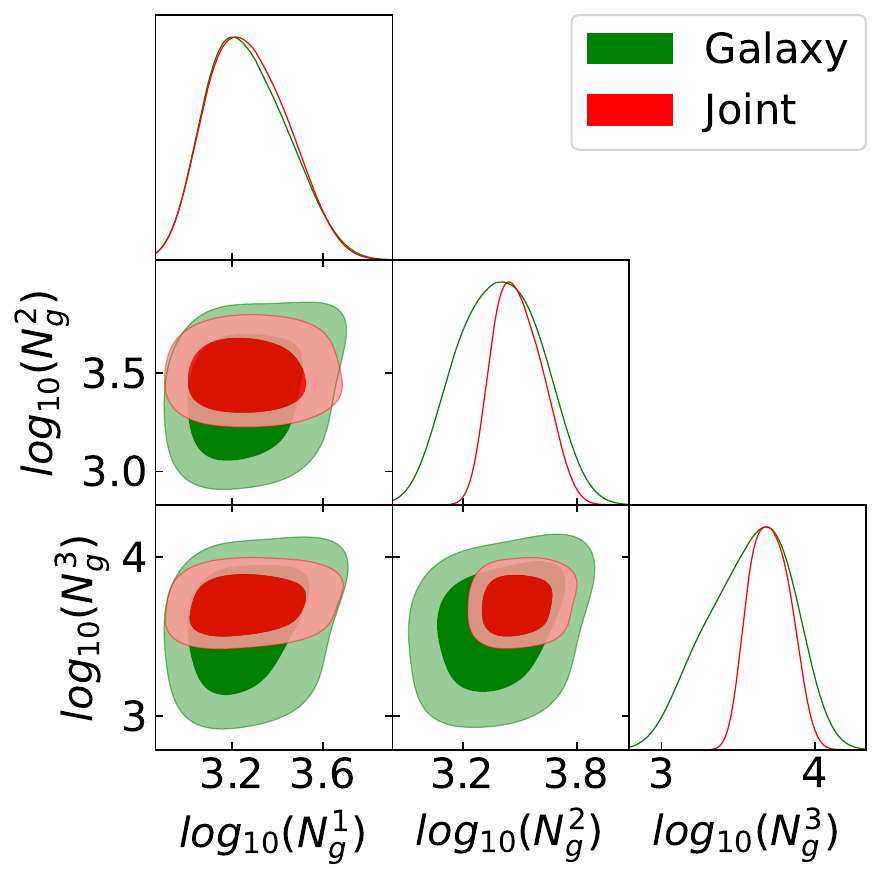}}
\subfigure{ \includegraphics[width=0.66\columnwidth]{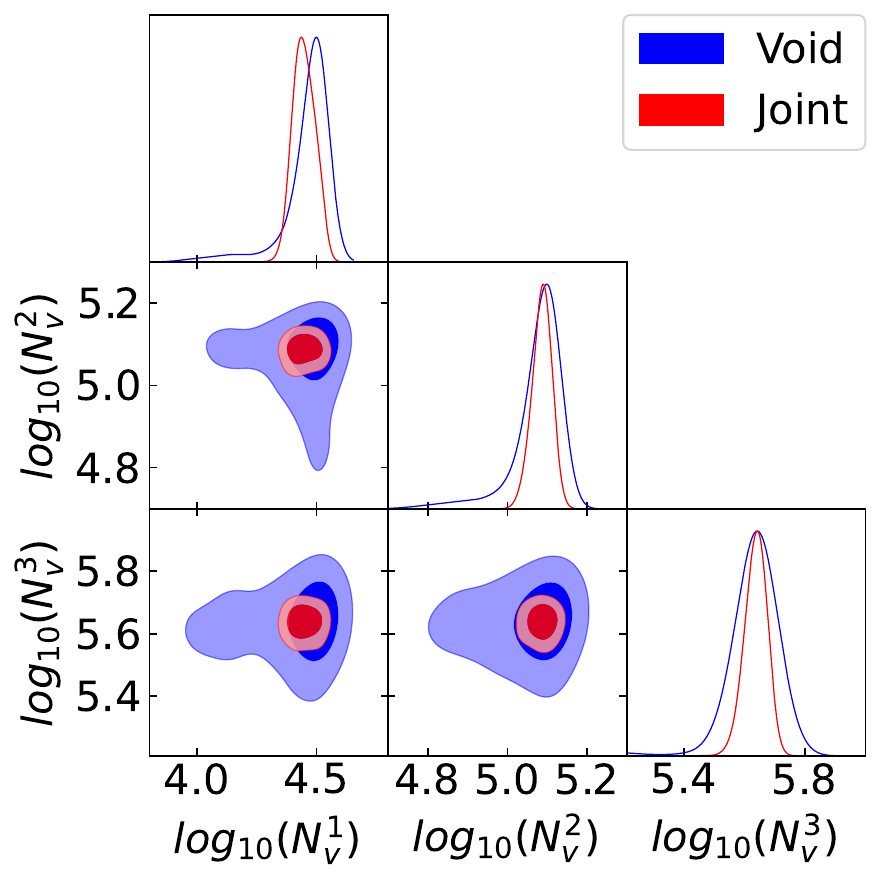}}
\subfigure{ \includegraphics[width=0.66\columnwidth]{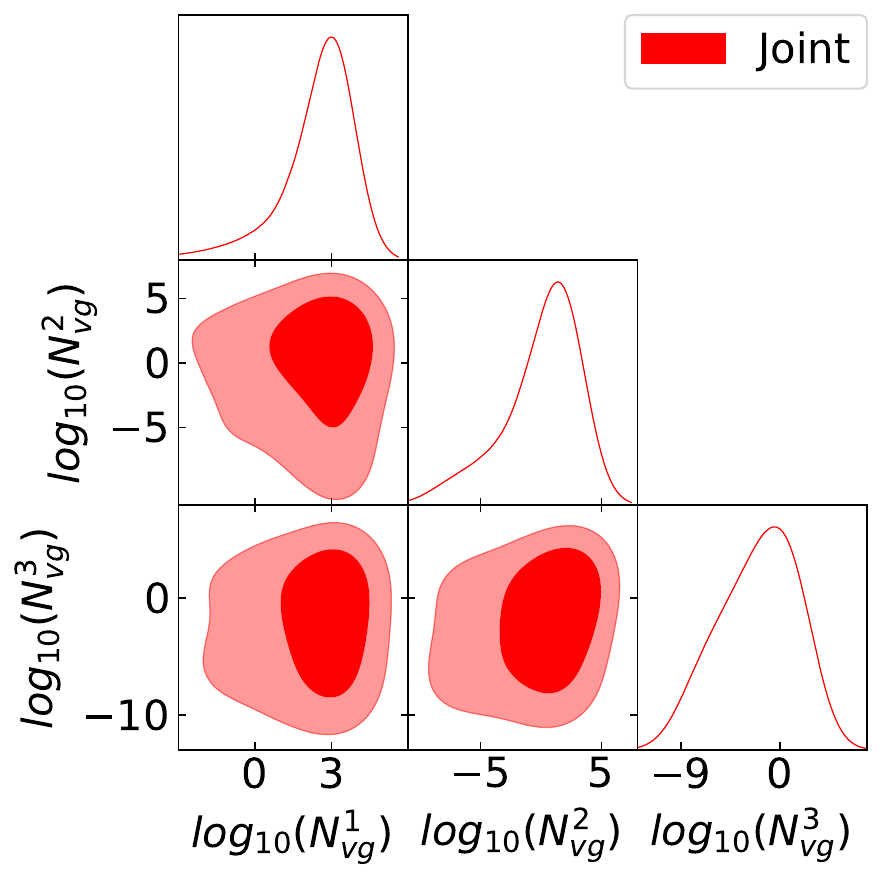}}
\caption{Contour maps at 68\% and 95\% CL and 1D PDFs of $N^i_{\rm g}$ (left panel), $N^i_{\rm v}$ (middle panel) and $N^i_{\rm vg}$ (right panel) at $z=0.3$, 0.6 and 0.9 derived from the galaxy power spectrum (green), void power spectrum (blue) and joint power spectra (red) in the simulation, respectively.}
\label{fig:mcmcnoise}
\end{figure*}

We show the constraint results of the galaxy bias in Figure \ref{fig:mcmcbias}. We obtain 
$b_{\rm g}^1=1.251_{-0.262}^{+0.302}$, $b_{\rm g}^2=1.575_{-0.315}^{+0.331}$, $b_{\rm g}^3=2.274_{-0.437}^{+0.494}$ in the joint fitting process. 
We can find the galaxy biases can be stringently constrained by the galaxy power spectra, and the joint constraint can provide a $\sim$15\% improvement on $b_{\rm g}^i$. This indicates that it is effective to include the void-galaxy cross power spectrum for studying the galaxy bias.

In Figure \ref{fig:mcmcnoise}, we show the constraint results of the noise terms in the galaxy, void and void-galaxy theoretical power spectra. We have 
${\rm log}_{10}(N_{\rm g}^1)=3.254_{-0.176}^{+0.204}$, ${\rm log}_{10}(N_{\rm g}^2)=3.476_{-0.125}^{+0.152}$, ${\rm log}_{10}(N_{\rm g}^3)=3.699_{-0.144}^{+0.144}$, ${\rm log}_{10}(N_{\rm v}^1)=4.443_{-0.042}^{+0.059}$, ${\rm log}_{10}(N_{\rm v}^2)=5.088_{-0.026}^{+0.022}$, ${\rm log}_{10}(N_{\rm v}^3)=5.640_{-0.039}^{+0.034}$, ${\rm log}_{10}(N_{\rm vg}^1)=2.993_{-1.680}^{+0.569}$, ${\rm log}_{10}(N_{\rm vg}^2)=1.069_{-4.213}^{+1.893}$ and ${\rm log}_{10}(N_{\rm vg}^3)=-1.369_{-4.861}^{+3.386}$ at the three redshifts in the joint fitting process. The joint constraints can provide a $\sim$50\% improvement on $N_{\rm g}$ at $z=0.6$ and 0.9 and $\sim$25\% improvement on $N_{\rm v}$ at $z=0.3,0.6$ and 0.9. We can find that the noise terms in the galaxy and void auto power spectrum increase with redshift, and $N_{\rm vg}$ in the void-galaxy cross power spectrum is relatively low and close to zero at high redshifts.

\bibliography{ps}
\bibliographystyle{aasjournal}

\end{document}